\documentclass[11pt]{article}
\usepackage{sint,macros,macros_static,cite}
\usepackage{amssymb,amsmath}
\usepackage{epsfig}
\begin{document}
\begin{titlepage}

\begin{flushright}
MS-TP-03-13\\
DESY 03-156\\
CERN-TH/2003-233\\
SFB/CPP-03-45
\end{flushright}

\vskip 0.75cm
\begin{center}
{\Large\bf 
Non-perturbative Heavy Quark Effective Theory\\[0.5ex] 
}
\end{center}
\vskip 1.0cm
\vbox{
\centerline{
\epsfxsize=2.5 true cm
\epsfbox{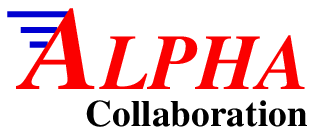}}
}
\vskip 1.0cm
\begin{center}
{\large
Jochen Heitger$^{\scriptscriptstyle a}$ and
Rainer Sommer$^{\scriptscriptstyle b,c}$
}
\vskip 1.0cm
$^{\scriptstyle a}$
Westf\"alische Wilhelms-Universit\"at M\"unster,
Institut f\"ur Theoretische Physik, \\
Wilhelm-Klemm-Strasse~9, D-48149 M\"unster, Germany
\vskip 2.5ex
$^{\scriptstyle b}$
Deutsches Elektronen-Synchrotron DESY, Zeuthen, \\
Platanenallee~6, D-15738 Zeuthen, Germany
\vskip 2.5ex
$^{\scriptstyle c}$
CERN, Theory Division, CH-1211 Geneva 23, Switzerland
\vskip 0.875cm
{\bf Abstract}
\vskip 0.7ex
\end{center}

We explain how to perform non-perturbative computations in HQET on the
lattice. In particular the problem of the subtraction of power-law
divergences is solved by a non-perturbative matching of HQET and QCD. 
As examples, we present a full calculation of the mass of the b-quark 
in the combined static and quenched approximation and outline an 
alternative way to obtain the B-meson decay constant at lowest order. 
Since no excessively large lattices are required, our strategy can also
be applied including dynamical fermions. 



\vskip 0.29cm
\vfill

\begin{center}
% \today 
October 2003
\end{center}

\eject
\vfill
\eject

\end{titlepage}

\section{Introduction \label{s_intro}}
The physics of the mixing and decays of B-mesons 
is essential for a determination of unknown
CKM-matrix elements and thus for our understanding
of the violation of CP-symmetry in Nature. It is also
still promising for the discovery of physics beyond
the standard model of particle physics.
Unfortunately, many of the experimental observations can only be related
to the standard model parameters if transition matrix
elements of the effective weak Hamiltonian are known. 
These matrix elements between hadron states 
are only computable in a fully non-perturbative
framework. They provide a strong motivation to study
B-physics in lattice QCD. However, as the mass of the b-quark
is larger than the affordable inverse lattice spacing
in Monte Carlo simulations of lattice QCD, this quark escapes a
direct treatment as a relativistic particle. Therefore, effective
theories for the b-quark are being developed and used to
compute the matrix elements in question~\cite{lat02:yamada,lat03:kronfeld}.

The first --- and very promising --- effective theory that 
was suggested is the 
Heavy Quark Effective Theory (HQET) \cite{stat:eichten,stat:eichhill1}.
Like others, it is afflicted by
a problem which remained unsolved so far: in general its
parameters (the coefficients of the terms in the Lagrangian)
themselves have to be determined non-perturbatively,
as briefly explained in \sect{s_power}. 
In other words, the theory
has to be renormalized non-perturbatively \cite{stat:MaMaSa}. 
This fact is simply due
to the mixing of operators of different dimensions in
the Lagrangian, requiring fine-tuning of their coefficients.
If they were determined only 
perturbatively (in the QCD coupling), the 
continuum limit of the theory would not exist. 

The issue is already present
in the determination of the b-quark mass in the static
approximation, i.e.~in the lowest order of the effective theory.
In \cite{lat01:rainer} a strategy was introduced and
successfully applied to this problem for the first time,
and a general framework for a non-perturbative renormalization
of HQET was sketched in \cite{lat02:rainer}. The basic idea,
illustrated in \fig{f_mbstrat}, is easily explained. 
%
%%%%%%%%%%%%%%%%%%%%%%%%%%%%%%%%%%%%%%%%%%%%%%%%%%%%%%%%%%%%%%%%%%%%
\newcommand{\ftext}[1]{\fbox{ {#1} }}
%
%% black and white
%
\newcommand{\cred}{}
\newcommand{\cblu}{}
\newcommand{\cmag}{}
\newcommand{\cgre}{}
\newcommand{\cbla}{}
\newcommand{\mgt}{\cmag}
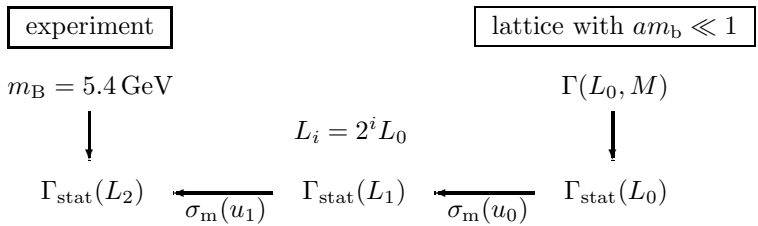
\begin{figure}[b]
% \vspace{2.5cm}
% \hspace{1.25cm}
\vspace{2.0cm}
\hspace{1.75cm}
\begin{picture}(8,20)(0,0)
\small
  \unitlength 0.4cm
  \put(2,6){\ftext{experiment}}            
  \put(17.5,6){\ftext{lattice with $a\mbeauty\ll 1$}} 
  \put(1.7,4){ $\mB=5.4\,\GeV$}    \put(20.1,4){ $\meff(L_0,M)$} 
  \linethickness{0.3mm}\cgre\put(4.7,3.4){\line(0,-1){1.5}}
  \linethickness{0.3mm}\cgre\put(4.7,1.8){\vector(0,-1){0.01}}
  \linethickness{0.3mm}\cgre\put(22.1,3.4){\line(0,-1){1.5}}
  \linethickness{0.3mm}\cgre\put(22.1,1.8){\vector(0,-1){0.01}}
  \linethickness{0.3mm}\cbla
  \put(2.8,0.5){ $\meffstat(L_2)$}
  \put(11.5,0.5){ $\meffstat(L_1)$}
  \put(20.2,0.5){ $\meffstat(L_0)$}
  \put(19.5,0.7){\line(-1,0){3.2}}
  \put(16.2,0.7){\vector(-1,0){0.01}}
  \put(10.8,0.7){\line(-1,0){3.2}}
  \put(7.5,0.7){\vector(-1,0){0.01}}
  \put(16.6,-0.1){$\sigmam(u_0)$}
  \put( 7.9,-0.1){$\sigmam(u_1)$}
  \put(11.5,2.5){\small $L_i = 2^i L_0$}
\end{picture}
% \vspace{-0.5cm}
% \vspace{4cm}
\caption{\label{f_mbstrat} \footnotesize
Relating experimental observables to properly renormalized HQET.
$\meffstat$ is a renormalized quantity in HQET and $\sigmam(\gbar^2(L))$
connects $\meffstat(L)$ and $\meffstat(2L)$; their exact definitions will
be given in the course of this paper. In the chosen example,
the experimental observable is the mass of the B-meson.}
\end{figure}
% \vspace{0.5cm}
%%%%%%%%%%%%%%%%%%%%%%%%%%%%%%%%%%%%%%%%%%%%%%%%%%%%%%%%%%%%%%%%%%%%

%
In a finite volume of linear
extent $L_0=\rmO(0.2\,\fm)$, one may realize lattices with
$a\mbeauty \ll 1$ such that 
the b-quark can be treated as a standard relativistic fermion. 
At the same time
the energy scale $1/L_0=\rmO(1\,\GeV)$ is still significantly below 
$\mbeauty$ and HQET applies quantitatively. 
Computing the same suitable observables
in both theories relates the parameters of HQET to those of QCD.
Then one moves, by an iterative procedure that we can still leave 
unspecified here, to larger and larger volumes and computes HQET 
observables.
This yields the connection to a physically large volume (of linear extent 
$\rmO(2\,\fm)$), where eventually the desired matrix elements are 
accessible.

Since in this way the parameters of HQET are determined from those 
of QCD, the predictive power of QCD is transfered to HQET.
In addition to solving the renormalization problem of the effective 
theory,
one also eliminates the usual need to determine more and
more parameters of the theory from experiment as the effective
theory is considered to higher and higher order. 

Although related, the strategy we propose here is not to be confused 
with the one for
the computation of the running of the coupling and 
renormalization group invariant matrix elements as 
first suggested by L\"uscher, Weisz and Wolff 
\cite{alpha:sigma} and then developed by the ALPHA Collaboration. 
We will discuss the difference in \sect{s_relation}.

In this paper we define the effective theory in detail, 
discussing in particular its renormalization properties 
(\sect{s_hqet}).
We then explain the matching between QCD and HQET 
(\sect{s_match}) as well as the finite-size strategy 
(\sect{s_finvol}) in the general case. 
\sect{s_exple} provides two examples of applications using the effective
theory at the lowest order in the inverse b-quark mass. 
The first one is the computation
of the quark mass, where numerical results illustrate
that indeed the power-law divergence can be subtracted 
non-perturbatively, retaining a very good precision for the final 
physical number. The second one, devoted to the B-meson decay constant,
has not yet been applied numerically but is a useful and {\em simple} 
example to help in understanding our method.
In \sect{s_uncert} we discuss the potential of our approach as well
as the expected uncertainty due to the use of a finite order in the
HQET expansion.

\section{HQET on the lattice \label{s_hqet}}
In this section we define the effective field theory for QCD containing
a heavy quark flavour in lattice regularization, starting from the
formal $\minv$--expansion of the classical theory. 
We drop all terms involving the heavy anti-quark fields as
they can be incorporated in complete analogy to those 
containing the heavy quark field $\heavy$ which we discuss
in detail.\footnote{For simplicity we drop higher-dimensional 
operators in the effective field theory which involve only 
light quark fields and the gluon field. These terms contribute at 
higher order in $\minv$.}
Renormalization properties are addressed but the proper choice of 
renormalization conditions and the physics content of the theory
is deferred to the next section.
\subsection{Definition of the effective theory}
We consider QCD on the lattice.
The explicit form of the gauge field and light fermion
action, $\Srel$, is not needed for our general
discussion, but for some of the following statements to hold,
an $\Oa$ improved formulation is required, e.g.~the one described in 
\cite{impr:pap1}.\footnote{$\Oa$ improvement means that the continuum limit
is reached with corrections of $\rmO(a^2)$.}
We denote the set of (bare) parameters of the theory 
with $\nf$ relativistic quarks by $\qcdpar$.
Apart from the gauge coupling, $g_0$, and the quark masses, it will 
in general also cover some improvement coefficients~\cite{impr:pap1}.

As has been explained by Eichten and Hill
\cite{stat:eichten,stat:eichhill1,stat:eichhill2},
an effective field theory for hadrons (at rest) containing 
$\nf-1$ light quarks and
one heavy quark (b-quark) with mass $m$ may be obtained by a formal 
$\minv$--expansion of the continuum QCD action and the fields, which appear
in the correlation function under study. 
The action of the heavy quark is written in terms of the four-component 
field $\heavy$ satisfying
\bes
  \label{e_constraint}
P_{+}\heavy=\heavy\,,\quad \heavyb P_{+}=\heavyb\,,\quad 
  P_{+}=\frac12(1+\gamma_0)\,.
\ees
Including terms up to the order $1/m^n$ in the expansion, 
the action, discretized on a Euclidean lattice, reads
\bes
  \label{e_shqet}
  \Shqet &=& a^4 \sum_x \Big\{ \Lstat(x) + \sum_{\nu=1}^n \lnu{\nu}(x)
                        \Big\} \,, 
\\
  \Lstat(x) &=& % \frac{1}{1+a\dmstat} 
               \heavyb(x)\,[\,\nabstar{0}+\dmstat\,]\,\heavy(x) \,, \qquad
\\[1.0ex]
   \lnu{\nu}(x) &=& \sum_i \lcoeff{\nu}{i}\opi{i}{\nu}(x)\,,
\ees 
where $\nabstar{\mu}$ denotes the backward lattice derivative,
$\dmstat$ has mass-dimension one, 
and the local composite fields $\opi{i}{\nu}$ have mass-dimension $4+\nu$. 
Indeed, this form is suggested by
a formal $\minv$--expansion at the classical level which yields\footnote{
A short derivation may e.g.~be found in \cite{nrqcd:first}.
}
\begin{alignat}{2}
&&
  \qquad  \dmstat &=0 \,,
\\
  \opi{1}{1} &= \heavyb\Big(-{\frac12\,\vecsigma\cdot\vecB}\Big)\heavy \,,&
  \qquad  \lcoeff{1}{1} &=\minv \,,
\\
  \opi{2}{1} &= \heavyb\Big(-{\frac12\,\vecD^2}\Big)\heavy \,,& 
  \qquad  \lcoeff{1}{2} &= \minv 
\end{alignat}
up to and including the order $\minv$. Here, $\vecB$
is a discretized version of the chromomagnetic field strength
and $\vecD^2$ a lattice version 
of the covariant Laplacian in three dimensions.
Note that a term $m\,\heavyb(x)\heavy(x)$ has been removed from the 
action, since it only corresponds to a universal energy shift 
of all states containing a heavy quark. Removing it makes explicit 
that the dynamics of heavy-light systems is independent of
the scale $m$ at lowest order of $\minv$. 

While the action is sufficient to obtain energy levels,
for many applications one is interested in (e.g.~electroweak transition 
matrix elements) it becomes necessary to also discuss 
correlation functions of composite fields. As an example we take the
time component of the axial current. In the effective theory
it is defined by an expansion similar to \eq{e_shqet},
\bes
  \Ahqet(x) &=& \sum_{\nu=0}^n \opa{}{\nu}(x)\,,
\\
  \label{e_StatAxial}
  \opa{}{0}(x)&=&\acoeff{0}{0} \Astat(x)\,, 
                 \quad \Astat(x)=\lightb(x)\gamma_0 \gamma_5\heavy(x)\,,
\\[0.5ex]
  \opa{}{\nu}(x) &=& \sum_i \acoeff{\nu}{i}\opa{i}{\nu}(x)\,,\quad\nu>0\,,
\ees
where a light quark field, $\light$, enters and $\opa{i}{\nu}$
is of dimension $3+\nu$.
One may then study for instance the correlator
(with $(\psibar_i\Gamma\psi_j)^{\dagger}\equiv
\psibar_j\gamma_0\Gamma^\dagger\gamma_0 \psi_i$)
\bes
  \label{e_caahqet}
    \caahqet(x_0) =  a^3\sum_{\vecx} 
          \Big\langle \Ahqet(x)\big(\Ahqet\big)^{\dagger}(0) \Big\rangle\,. 
\ees
At the classical level the fields are given by
\bes
     \acoeff{0}{0}=1 \,,\quad 
     \opa{1}{1} = \lightb\gamma_j\gamma_5 \ola{D}_j\heavy \,,\quad
     \acoeff{1}{1} =\minv \,.
\ees
In general, i.e.~at the quantum level, 
expectation values are defined by a path integral 
\bes \label{e_pathi}
  \langle \op{} \rangle &=& \frac1Z \int D[\varphi]\, \op{}[\varphi]\,
                            \rme^{-(\Srel + \Shqet)} \,,\\
                      Z &=& \int D[\varphi]\, \rme^{-(\Srel + \Shqet)} \,,
\ees
over all fields with the standard measure, denoted here by $D[\varphi]$.
An important ingredient in the definition of the
effective field theory is that it is understood throughout 
that the {\em integrand} of the path integral is expanded 
{\em in a power series} in $\minv$, with power counting according to
\bes
  \label{e_counting}
  \lcoeff{\nu}{i} = \rmO\left(1/m^\nu\right) \,, \quad
  \acoeff{\nu}{i} = \rmO\left(1/m^\nu\right) \,.
\ees
In other words one replaces
\begin{multline} \label{e_expact}
 \exp\left\{-(\Srel + \Shqet)\right\} =  
 \exp\bigg\{-\Big(\Srel + a^4 \sum_x \Lstat(x)\Big)\bigg\} \,\times \\
            \bigg\{ 1 -  a^4 \sum_x \lnu{1}(x) 
                        + \frac12\,\Big[\,a^4 \sum_x \lnu{1}(x)\,\Big]^2
                        -  a^4 \sum_x \lnu{2}(x) + \ldots 
            \bigg\}                        
\end{multline}
in \eq{e_pathi}.
The $\minv$--terms then appear only as insertions of local operators
${\mathcal O}_{i}^{(\nu)}(x)$ and $\opa{i}{\nu}(x)$ into 
correlation functions, and the true path integral average is taken
with respect to the action in the static approximation for the heavy quark,  
$S=\Srel+a^4 \sum_x \Lstat(x)$. 

Power counting leads us to expect that the static theory is  
{\em renormalizable}, requiring a finite number of parameters 
to be fixed to obtain a continuum limit. Indeed explicit perturbative
\cite{stat:eichhill_za,stat:boucaud_za,stat:MaMaSa,zastat:pap1,zastat:pap2} 
as well as non-perturbative \cite{zastat:pap3} computations
support that this is a genuine property of the static effective theory.
Would one keep one of the 
$\minv$--terms in the exponent, as it is done in NRQCD,
renormalizability would be lost and most of what we
are concluding in this paper would not be true.

We are still left to discuss the renormalization of expectation 
values of the type (\ref{e_pathi}) after inserting the expansion
(\ref{e_expact}). This is just the problem of renormalizing 
correlation functions of local composite operators in
the  static effective theory.
Power counting immediately leads to the conclusion:
once {\it all local operators}, whose dimensions do not exceed the one 
of the highest-dimensional operator (i.e.~$\nu \leq n$) and which have the 
proper symmetries, are included, their coefficients may be chosen such that 
all expectation values have a continuum limit 
(see e.g.~Ref.~\cite{Renorm:collins}).
Of course, both the operators $\opi{i}{\nu}(x)$ in the action
and the ones in the effective operators such as $\opa{i}{\nu}(x)$
have to be included.
One may worry that due to the sums over all space-time points in 
\eq{e_expact} contact terms appear, which lead to additional
singularities.
However, just like in the case of $\Oa$ improvement discussed
thoroughly in \cite{impr:pap1}, the terms needed to
remove these singularities are already present once 
all local operators with the appropriate dimensions are included. 

The effective theory is now defined in terms of the set of parameters,
\bes
 \hqetpar \equiv \{c_k\} =%\{m_\up,m_\down,m_\strange,m_\charm\} 
         \qcdparred \cup \{\dmstat\} \cup \{\lcoeff{\nu}{i}\} \cup 
         \{\acoeff{\nu}{j}\} 
         \cup \ldots \,, \quad c_1\equiv g_0^2 \,.
\ees
The ellipses
allow for coefficients of further 
composite operators which will be needed when their 
correlation functions are considered.
For the continuum limit of this effective
theory to exist, the parameters $\{c_k, k>1\}$
have to be chosen properly as a function of $g_0^2$. 
Note that in the notation used here, the 
renormalization  of the effective composite fields 
is included in the set of generalized coupling constants, $\hqetpar$.
E.g.~at lowest order
in $\minv$, the coefficient $\acoeff{0}{0}\equiv\zastat$ 
is the renormalization constant of the static axial current 
\cite{zastat:pap1}.
 
A few more remarks are in order.
\bi
 \item  Once the proper degrees of freedom, namely the field 
	$\heavy$, have been identified,
        the terms in the effective field theory 
        are organized just by their mass-dimension.
        The expectation that the effective field theory has a continuum
        limit (is non-perturbatively renormalizable) is thus nothing 
        but the usual expectation that composite operators mix only with
        operators of the same and lower dimension.
 \item  The same argumentation is also the basis of Symanzik's 
        discussion of cutoff effects of lattice theories and their
        removal order by order in $a$ 
        \cite{impr:Sym1,impr:Sym2,impr:onshell,impr:pap1}. 
        An important consequence is
        that in general the $\minv$--expansion and the $a$--expansion are 
        not independent but have to be considered as one expansion in 
	terms of the dimension of the 
        local operators. If we imagine to start with a theory with a 
	set of operators identified by the formal continuum 
        $\minv$--expansion, these operators will for instance mix under 
        renormalization with operators of the same and lower dimension, 
        which are allowed by the lattice symmetries but not by the 
        continuum symmetries and which would therefore not be in the set 
        of the  operators one started with. 
        To avoid this, one has to start immediately
	with the full set of operators of a given dimension, restricted
	only by the lattice symmetries. In other words we have to
	count $a = \rmO(\minv)$.
        This means also that $\Srel$ has to be $\Oa$ improved
        to go to order $\minv$.\footnote{
        It may be possible to go to higher order in $a$ than in $\minv$,
        when symmetries restrict the allowed mixings.
        An example is provided by $\Oa$ improvement of the static effective
        theory~\cite{zastat:pap1}. 
        Ways to extend this to higher orders in $\minv$ probably exist but we
        have not investigated this question systematically.}
 \item  Of course, symmetries restrict the terms that have to
        be taken into account. In general, out of the space-time
        symmetries we only have the 3--dimensional
        cubic group instead of the 4--dimensional hypercubic group.
        At the lowest order in $\minv$ there are additional symmetries:
        heavy quark spin-symmetry~\cite{stat:symm1} 
        and the local conservation of heavy quark number, which simplify
        $\Oa$ improvement (see Sect.~2.2 of \cite{zastat:pap1}). 
 \item  Furthermore it is convenient to formulate the effective
        theory only on-shell, i.e.~for low energies as well as
        for correlation functions at physical separations. Then the
        argumentation of \cite{impr:pap1} can be taken over literally
        to show that the equations of motion 
        (derived from the lowest-order action) can be used to reduce the
        set of operators $\opi{i}{\nu}(x),\opa{i}{\nu}(x),...\,$.
        Following the same reference, operators obtained by multiplying 
        those of dimension $d$
        by a light quark mass are to be counted as
        separate operators of dimension $d+1$.
 \item  Finally note that after using \eq{e_expact}, the determinant
        arising from the static quark action is just an irrelevant constant.
        In principle, loop effects of the heavy quark are still present in 
        the coefficients $c_k$.
\ei 
\subsection{Power divergences and non-perturbative 
            renormalization \label{s_power}}
The mixing of operators differing in dimensions by $p$
translates into coefficients diverging (when $p>0$)
as $a^{-p}$. In the present context it actually has been checked in 
perturbation theory that these mixings are not forbidden 
by some accidental symmetry \cite{stat:MaMaSa}. 
Due to such power divergences, 
perturbation theory is not sufficient to determine the 
coefficients $c_k$.
An estimate of order $g_0^{2l}$
would leave a perturbative remainder
\bes
  \Delta c_k \sim g_0^{2(l+1)}\,a^{-p} \;\sim\; a^{-p}\,
     [\ln(a\Lambda)]^{-(l+1)} 
  \;\;\toas{a\to0} \;\;\;  \infty \,\quad % \text{for}\;\nu < n\,,
\ees 
with $\Lambda$ the QCD $\Lambda$--parameter.
This means that the continuum limit does not exist if the coefficients 
are determined only perturbatively.

Hence we conclude that a non-perturbative method is needed to determine
(at least some of)
the parameters $\{c_k\}$. Such a method will be introduced in the following
two sections.

\section{Matching of HQET and QCD \label{s_match}}
By QCD we denote the theory 
including a relativistic heavy quark, the b-quark,
while with HQET we mean the theory where this quark
is incorporated with the action defined in the previous section.
The latter is an approximation to QCD  
when the coefficients $\hqetpar =\{c_k\}$ 
are chosen correctly. Then
we expect 
\bes
\label{e_equiv}
  \Phihqet(M) = \Phiqcd(M) + \rmO\left(1/M^{n+1}\right)
\ees
for properly chosen observables, $\Phiqcd$, in QCD and
their counterparts, $\Phihqet$, in the effective theory.
Amongst the many dependencies of $\Phiqcd$ we have
indicated only the one on the  
heavy quark mass. To be free of any renormalization
scheme dependence,  we choose the
renormalization group invariant (RGI) quark mass 
denoted by $M$ \cite{mbar:pap1}. In order for \eq{e_equiv} to hold,
all other scales appearing
in $\Phiqcd$ are assumed to be small compared to $M$. 
Choosing as a typical low-energy reference scale of
QCD the energy scale $r_0^{-1}$($\approx 400\,\MeV$) \cite{pot:r0}, 
defined in terms of the
QCD force between static quarks, the combination
$r_0 M$ has to be large.
Thus the symbol $\rmO(1/M^n)$ is a short hand for 
$\rmO(1/[r_0 M]^n)$.

To give a {\em simple example} for a quantity $\Phiqcd$, one could take
$\Phiqcd = \caa$, where
\bes \label{e_caa}
  \caa(x_0) = \za^2 a^3\sum_{\vecx} \Big\langle A_0(x)  (A_0)^{\dagger}(0) 
              \Big\rangle
\ees
with the heavy-light axial current in QCD, 
$A_\mu=\lightb\gamma_\mu\gamma_5\psi_\beauty$, and $\za$ ensuring 
the natural normalization of the current consistent with 
current algebra~\cite{Boch,impr:pap4}.   
Then \eq{e_equiv} is valid for $\Phihqet = \rme^{-m x_0}\caahqet(x_0)$
with $\caahqet(x_0)$ from \eq{e_caahqet}
and in the region  $1/x_0 \ll M$. With the latter, kinematical, condition 
one takes care that the correlation functions are dominated by states
with energies (the heavy quark mass being subtracted) small compared to $M$.
Furthermore, the factor $\rme^{-m x_0}$ accounts for the mass term 
that had been removed from the effective theory Lagrangian as already
mentioned after \eq{e_shqet}.
Which mass $m$ is to be taken here, depends on the convention 
used to define $\dmstat$. As will be explained further 
in \sect{s_strat},
at each order in $\minv$, the combination 
$m + \frac{1}{a} \ln(1+a\dmstat)$ 
is uniquely fixed by the matching of HQET and QCD. We
emphasize again that the same mass $m$ enters all 
correlation functions involving one heavy quark.

Let us now come to the main problem: determining the parameters in
the effective theory such that this equivalence between HQET and QCD is 
true. First of all assume that the parameters of QCD have been
fixed by requiring a set of observables, e.g.~a set of hadron masses, 
to agree with experiment. 
It is then sufficient to impose  
\bes
  \label{e_match}
  \Phihqet_k(M) = \Phiqcd_k(M)\,, \quad k=1,\ldots,\Nn\,,
\ees
to determine all parameters
$\{c_k\,,k=1,\ldots,\Nn\}$ in the effective theory. 
Observables
used originally to fix the parameters of QCD may be amongst these
$\Phiqcd_k$.
The matching conditions, \eq{e_match}, define the set 
$\{c_k\}$ for any value of the lattice spacing 
(precisely speaking, for any value of $a/r_0$). 

In principle, each $\Phihqet_k$ could be determined from a 
physical, experimentally accessible observable. 
However, this
would reduce the predictive power of the effective theory since
it contains more parameters than QCD.
Particularly for increasing the order $n$ of the $\minv$--expansion 
we then would need to use more and more experimental observables.

To preserve the predictability of the theory, we may instead insert some 
quantities $\Phiqcd_k(M)$ computed in the continuum limit
of lattice QCD. 
This of course demands
to treat the heavy quark as a relativistic particle
on the lattice, seemingly in contradiction
to the very reason to consider the 
effective theory: small enough lattice spacings 
to do this are very difficult to reach. An additional 
ingredient is thus necessary to make the idea practicable.
It will be explained in the following section. At this stage  
the important point is that there are no theoretical 
obstacles to a non-perturbative matching. 
We end this section with some comments on details of the 
general matching procedure.
\bi
\item   The observables $\Phi$ are assumed to be renormalized.
        \Eq{e_match} is, however, used to fix the bare parameters in the 
        action --- for each value of $g_0^2$.
\item   When one increases the order $n$ in the expansion, new
        quantities $\Phi_k$ have to be added, and at the same time,
        the parameters of the lower-order Lagrangian,
        $c_i$, $i \leq N_{n-1}$, will change in general. This change
        is due to mixing of the operators and may thus be sizeable.
\item   It is convenient to take the continuum limit\footnote{
                Or alternatively, work in a sufficiently improved
                lattice theory  and at a small value of 
                the lattice spacing.}
        of $\Phiqcd_k$ before imposing \eq{e_match}. If one decides
        not to do this, 
        the lattice spacings on both sides of \eq{e_match} should
        be scaled together in order to reach the continuum limit in
        the effective theory. 
\item   As mentioned already in the previous section, the terms
        necessary for Symanzik improvement are taken into account 
	automatically, 
        namely some of the equations (\ref{e_match}) may be interpreted
	as improvement conditions. Working up to the order $n$,
        the resulting lattice HQET is correct up to 
        \bes
           \text{error terms} = \rmO\left((\minv)^{n+1}\right) = 
                                \rmO\left(M^{-(n+1)}(aM)^{k}\right)\,,
           \quad k=0,1,\ldots,n+1 \,.
        \ees
        Higher-order terms in $1/M$ have parametrically
        larger lattice spacing errors. For example,
	a treatment of the theory including the next-to-leading 
	operators will give us the $(1/M)^0$--terms with 
        $\rmO(a^2)$ errors and the linear $1/M$--corrections with $\Oa$ 
        uncertainties.
	Additional work would be necessary to suppress the 
	discretization effects in the $1/M$--terms to $\rmO(a^2)$.
\item   There is a close analogy of our proposed matching procedure
        to what is done when the low-energy constants of the 
        chiral effective Lagrangian~\cite{chir:GaLe1} are determined using
        lattice QCD. An important difference is, however, that
        the chiral expansion can be worked out analytically 
        while here we still have to evaluate the resulting
        theory by Monte Carlo. The reason is that strong interactions remain;
        the lowest-order theory, the static approximation, is non-trivial.
\ei

\section{The r\^ole of finite volume \label{s_finvol}}
From the theoretical point of view, the matching described in the 
previous section is sufficient. 
However, we should take into consideration what
can be done in a numerical computation. To give a concrete
example, let us assume that 
\bes
(L/a)^3\times T/a \leq 32^3 \times 64
\ees
lattices can be simulated,
numbers which are realistic for present computations in the
quenched approximation, but too large for full QCD. We further assume
that we deal with quantities which have negligible finite size effects
when 
\bes
 L \ge 2\,\fm\,.
\ees
Then the smallest lattice spacing reachable is $a\approx0.06\,\fm$,
and this number will not be very different if the above assumptions
are modified within reasonable limits.
While such a lattice spacing is small
enough to perform computations for charm quarks~\cite{mbar:charm1,fp:fds1}, 
the 
subtracted bare mass of the b-quark is about $a\mq \approx 1$.
In this situation lattice artifacts are expected to be very large
and it is impossible to obtain the r.h.s.~of \eq{e_match}. 

The situation becomes quite different when one considers observables
$\Phi_k$ defined in finite volume with $L$ considerably smaller 
than $2\,\fm$ and uses the generally accepted --- and also
much tested --- assumption that
{\em both QCD and HQET are applicable in a finite volume and the
parameters in the Lagrangians are independent of the volume}. 
\subsection{Matching in finite volume}
Instead of \eq{e_match} we now consider (remember that $\Nn$ is the number
of parameters in the effective theory):
\bes
  \label{e_match_l}
  \Phihqet_k(L,M) = \Phiqcd_k(L,M)\,, \quad k=1,\ldots,\Nn\,.
\ees
This will allow us to have much smaller lattice spacings on the 
r.h.s.~in order to eventually approach the continuum limit.
A typical choice is $L=L_0 \approx 0.2\,\fm$. 
As has been shown in the preliminary report of our 
work \cite{lat01:rainer}, \eq{e_match_l} can be evaluated very precisely 
for suitably selected quantities $\Phiqcd_k$ and the continuum limit can 
actually be taken. 

Concerning the l.h.s., we have to take into account that
$a$--effects will certainly be significant when the resolution $a/L$ of the 
finite space-time is too coarse. Hence the lattice spacings where 
the bare parameters $\{c_k(g_0)\}$ can be determined
are $a=\rmO(0.02\,\fm)=\rmO(L_0/10)$. 
For such values of $a$, the
computation of the physical observables in the infinite-volume
theory ($L \approx 2\,\fm$ in practice) would again be impracticable,
because lattices with too many points $(L/a)^4$ would be required.
Therefore, a further step is necessary to make larger lattice spacings 
and thereby larger physical volumes available in the effective theory.
\subsection{Finite-size scaling}
Also for this step a well-defined procedure is easily found. 
First assume that all observables $\Phihqet_k(L,M)$ have been made
dimensionless by multiplication with appropriate 
powers of $L$. Next we define step scaling 
functions~\cite{alpha:sigma}, $F_k$, by
\bes
\label{e_ssfcont}
   \Phihqet_k(sL,M)= 
   F_k\left(\big\{\Phihqet_j(L,M)\,,\,j=1,\ldots,\Nn\big\}\right) \,,
   \quad k=1,\ldots,\Nn \,,
\ees
where typically one uses scale changes of $s=2$.
These dimensionless functions describe the change of 
the complete set of observables $\{\Phihqet_k\}$ under a scaling of
$L\to sL$, and we briefly sketch how they can be computed.
One selects a lattice with a certain resolution 
$a/L$. The specification of $\Phihqet_j(L,M)$, $j=1,\ldots,\Nn$, then 
fixes all (bare) parameters of the theory. The l.h.s.~of \eq{e_ssfcont} 
is now computed,  keeping the bare parameters fixed
while changing $L/a \to L'/a = sL/a$. Repeating this for a few values 
of $a/L$, the continuum limit of $F_k$ can be obtained by an 
extrapolation $a/L\rightarrow 0$.

An important practical detail is to choose
the various quantities $\Phihqet_k$ such that each $F_k$ depends only on
a few $\Phihqet_j$ and the bare parameters $c_k$ can be
determined rather independently from each other. 
For instance, it is natural to identify the running Schr\"odinger 
functional coupling $\gbar^2(L)$ \cite{alpha:sigma,alpha:SU3} with 
$\Phihqet_1(L,M)$ and to keep all of the light quark masses zero in these 
steps. 
In this way $g_0^2$ and the (light) bare quark masses are 
fixed independently of the parameters 
$\dmstat,\lcoeff{\nu}{i},\acoeff{\nu}{i},\ldots$ coming from the heavy 
sector.
In the quenched approximation or with two dynamical quarks, the set of 
bare parameters specifying the relativistic sector, $\qcdparred$, can then 
be taken over from \cite{alpha:SU3,mbar:pap1,alpha:letter,lat02:michele} 
without change.

A few steps --- may be two --- are necessary
to reach a value of $L = \rmO(1\,\fm)$, where at the same time contact can
be made with resolutions $a/L$ that are affordable to accommodate the 
suitable observables on a physically large lattice to realize the original
matching condition, \eq{e_match}.
(In our first example, \sect{s_mb}, this r\^{o}le will be played by the 
B-meson with its mass as the physical input.)
We note that in principle the size of $L_0$ is rather arbitrary,
but the following consideration is important. 
We are matching at a finite value of $\minv$ and a finite order
$n$. Thus the final results will depend on which quantities have been used
to perform the matching. If one chooses quantities
with kinematics where the $\minv$--expansion is not accurate
(or even not applicable), this will translate into badly
determined parameters in the effective Lagrangian and large final 
truncation errors. For this reason, $L_0$ has to be chosen such that 
the $\minv$--expansion is applicable which means
\bes
   1/L_0  \ll m\,,
\ees
and $L_0$ cannot be too small. From these considerations it
appears that $L_0\approx 0.2\,\fm\, - \,0.4\,\fm$ is a good choice.  
\subsection{Evaluation of the physical observables in the effective 
            theory}
Physical observables usually have to be computed in large
volume which, for practical reasons, means at
lattice spacings around $1/20\,\fm$ to $1/10\,\fm$. 
In this region the bare parameters of the effective theory 
are determined as follows.

One chooses a suitable $K$ such that  
\bes
   L_K = s^K L_0 \approx 1\,\fm\,.
\ees
Iterated applications of \eq{e_ssfcont} give rise to recursion relations,
the solutions of which determine quantities $V_k\equiv\Phihqet_k(L_K,M)$
in the larger volume of extent $L_K$.
Next, regarding $\Phihqet_k(L_K,M)=V_k$ as a requirement while setting the 
number of lattice points to $L_K/a=\rmO(10)$ just fixes the bare 
parameters $\hqetpar$.
These bare parameters are then known at values of the lattice spacing,
where the computation of correlation functions in large volume is 
possible in the effective theory and masses and matrix elements can 
be extracted from their large-time behaviour.

Note that in the notation used here also the 
renormalization constants of the composite operators appearing
in the correlation functions are
amongst the ``bare parameters''. All quantities are thus renormalized 
entirely non-perturbatively.\footnote{This represents
an advantage in comparison to
\cite{zastat:pap3} where a last step using perturbation theory 
was necessary to get to 
the ``matching scheme'' \cite{lat02:rainer,zastat:pap3}, 
which here we achieve by virtue of \eq{e_match}.}
One may still wonder how $M$ itself is fixed. The answer to this
question is provided by the first of the two examples, which we 
will use now to illustrate the general strategy.

\section{Examples\label{s_exple}}
In this section we supply two applications of our non-perturbative 
matching strategy of HQET and QCD that up to now was formulated in 
rather general terms:
a full calculation of the b-quark mass in combined static and quenched 
approximations (\sect{s_mb}) and a proposal for a non-perturbative 
determination of multiplicatively renormalized matrix elements of the 
static-light axial current, which is different in spirit from 
Ref.~\cite{zastat:pap3} and still awaits a numerical investigation.
\subsection{The b-quark mass at lowest order \label{s_mb}}
Several determinations of the mass of the b-quark, which use 
the static approximation on the lattice (HQET to order $(\minv)^0$), 
have been published~\cite{lat01:ryan}. They all
rely on a perturbative estimate 
of $\dmstat$ \cite{mbstat:MaSa,mb:dmstat_Direnzo,mb:dmstat_Trottier} and 
suffer from a power-law divergence due to the mixing of
$\heavyb D_0 \heavy$ and $\heavyb \heavy$ as discussed in 
\sect{s_power}. Their precision is thus limited by the 
fact that a continuum limit can not be taken, and it is difficult
to estimate the associated uncertainty. We here explain
how a entirely non-perturbative computation can be done and will
also give a first result, which can easily be improved in 
precision in the near future. This step is also a prerequisite
to perform the matching of other quantities such as the axial 
current, since generically the quark mass enters in the matching step
(\ref{e_match_l}) for all $\Phi_k$.
\subsubsection{Strategy and basic formula \label{s_strat}}
As indicated already in \sect{s_finvol}, given a resolution $a/L$,
we fix $g_0^2$ such that the finite-volume running coupling of 
Ref.~\cite{alpha:SU3} takes a certain value. 
Furthermore we set the light quark masses to zero (with one exception
which will be discussed). 
In the language of \sect{s_finvol} we have 
\bes \label{e_condgbar}
  \Phihqet_1 &=& \gbar^2(L)\,,\\
  \label{e_condm}
  \Phihqet_{k+1} &=& m^{\rm PCAC}_{k} \,\,\,=\,\,\,0\,, 
  \quad k=1,\ldots,\nf-1\,,
\ees
in terms of the PCAC masses of the light flavour number $k$, 
$m^{\rm PCAC}_{k}$, and the running coupling $\gbar^2(L)$ in the
Schr\"odinger functional (SF) scheme \cite{alpha:SU3}. 
Other choices are possible, but the above is convenient in view of
the present numerical knowledge \cite{mbar:pap1,alpha:letter,lat02:michele}.
The box length $L$ is then parametrized through $\gbar^2(L)$. 
A very useful feature of eqs.~(\ref{e_condgbar}) and (\ref{e_condm})
is that they do not involve the heavy field at all and 
determine the bare coupling and quark masses
independently of the heavy sector; in particular these conditions 
are independent of the order $n$ of the expansion.  

In this section we are only concerned with energies and remain 
at lowest order in $\minv$. 
The only additional parameter in the Lagrangian to be fixed is $a\dmstat$, 
i.e.~one more condition corresponding to $k=\nf+1$ in
\eq{e_match_l} is needed. We start from a time-slice correlation 
function projected onto  spatial momentum zero 
containing one heavy quark (such as 
$\caa$, \eq{e_caa}). Denoting it generically as $C(x_0)$, in the 
logarithmic derivative
\bes \label{e_defgam}
  \meff = \frac{1}{2a} \ln\Big[\,C(x_0-a)/C(x_0+a)\,\Big] \qquad 
  \left(\mbox{$x_0/L$ fixed}\right) 
\ees
all multiplicative renormalization factors of $C(x_0)$ cancel. 
Below we shall use $x_0/L = 1/2$, but other choices are possible.
Replacing the fields in the correlation function
$C(x_0)$ by the corresponding effective fields defines $C^{\rm HQET}(x_0)$
in the effective theory. 
In the static approximation, its logarithmic derivative, $\meffstat$,
built as in \eq{e_defgam},
depends on $\dmstat$ in the simple form
\bes  \label{e_additive}
  \meffstat = \left. \meffstat\right|_{\dmstat=0} 
               + \frac{1}{a} \ln(1+a\dmstat)\,,
\ees
as is easily seen from the explicit form of the static quark propagator.
In large volume, which due to $x_0/L=\mbox{constant}$ also means large 
Euclidean time, $\meff=\meff(L,M)$ will turn into the mass of a 
b-hadron, say $\mb$. It is now obvious that
\bes
  \Phiqcd_{\nf+1}(L,M)&\equiv& L\,\meff\,,\quad   \\
 \Phihqet_{\nf+1}(L,M)&\equiv& L\,(\meffstat + m) \\ 
                      &   =  & L\left(\left. \meffstat\right|_{\dmstat=0}
                               + \mhbare\right)\,,\quad
                               \mhbare = m + \frac{1}{a} \ln(1+a\dmstat)\,, 
                               \nonumber
\ees
are sensible assignments to fix the combination $\mhbare$ via requiring
\bes \label{e_m1}
  \Phiqcd_{\nf+1}(L,M) = \Phihqet_{\nf+1}(L,M)\,.
\ees
Since $\dmstat$ and $m$ always appear in the combination $\mhbare$,
they may not be fixed separately, unless one arbitrarily defines 
$\dmstat$ by an additional condition.

Due to \eq{e_additive}, the step scaling function
\bes \label{e_sigmam}
  \sigmam\left(\gbar^2(L)\right) \equiv 
  2L\, [\,\meffstat(2L,M) - \meffstat(L,M)\,] 
\ees
is independent of $\mhbare$ and therefore also independent
of $M$; at lowest order in
$\minv$, energy differences in the effective theory do not depend
on the heavy quark mass. The step scaling function (\ref{e_sigmam}) is 
thus a particularly simple realization of \eq{e_ssfcont}. Together with 
the one for the running coupling~\cite{alpha:sigma,alpha:SU3},
\bes \label{e_sigma}
  \sigma(u) = \left. \gbar^2(2L)\right|_{\gbar^2(L)=u}\,,
\ees
it defines the sequence
\bes
  u_0 = \gbar^2(L_0)\,,
  & \quad & 
  w_0 = \left.L\,\meffstat\right|_{L=L_0}\,, 
  \\
  u_{k+1} = \sigma(u_k)\,, 
  & \quad & 
  w_{k+1} = 2 w_k + \sigmam(u_k)\,, 
  \label{e_wk}
\ees
which is easily seen to relate $\meffstat(L_K,M)$, with $L_K=2^K L_0$, 
to $\meffstat(L_0,M)$ when the sequence $u_{0},\ldots,u_{K-1}$ is known:
\bes \label{e_m2}
 L_0\meffstat(L_K,M) = L_0\meffstat(L_0,M) 
                       + \sum_{k=0}^{K-1} 2^{-(k+1)}\sigmam(u_k)\,. 
\ees
Suitable choices for $u_0$ and $K$ then 
allow to arrive at $L_K=\rmO(1\,\fm)$.

Finally one considers the energy $\Estat$ of a B-meson in static 
approximation, given for example by
\bes
  \caahqet(x_0) \,\simas{x_0\to\infty}\, A \exp(-x_0 \Estat) \qquad
  \left(\mbox{$L$ large}\right)\,.
\ees
The energy difference
\bes \label{e_m3}
 \Delta E = \Estat - \meffstat(L_K,M)
\ees
can be computed with one and the same lattice spacing 
(i.e.~at the same bare parameters) for the two different terms on the 
r.h.s., but of course with different $L$.
Combining \eq{e_m1} imposed in small volume ($L=L_0$) with eqs.~(\ref{e_m2}) 
and (\ref{e_m3}) to eliminate $\mhbare$ in $\mb=\Estat+\mhbare$ 
(holding in the large-volume limit), we arrive at the basic equation
\bes \label{e_mall}
    L_0 \mb = L_0 \meff(L_0,\Mbeauty) + 
    \sum_{k=0}^{K-1}2^{-(k+1)} \sigmam(u_k) + L_0 \Delta E \,.
\ees
It relates the mass of the B-meson to a quantity 
$\meff(L_0,\Mbeauty)$, computable in lattice QCD with a relativistic
b-quark, and the energy differences $\Delta E$ and $\sigmam$ which are
both defined and computable in the effective theory. All quantities
on the r.h.s.~may be evaluated in the continuum limit. Note that 
all of the terms in \eq{e_mall} are independent of 
$\mhbare$ (because the unknown $\mhbare$, and thereby $\dmstat$ too, drop 
out in the differences), although logically \eq{e_m1} has been used to fix 
it non-perturbatively.
Our strategy has also been presented in a somewhat different
way, which is closer in spirit and notation to standard HQET applications 
(but less rigorous), in \cite{reviews:laurent}.

\Eq{e_mall} may be looked at in two different ways. 
Given the RGI mass
of the b-quark, $\Mbeauty$, \eq{e_mall} 
provides a way to compute the mass of the 
B-meson. It is more interesting to turn this around: 
taking $\mb$ from experiment and evaluating (in lattice
QCD)  $\meff(L_0,M)$ as a function of $M$, this equation
may be solved for $\Mbeauty$. Implicitly the bare parameter $\mhbare$
is thus fixed non-perturbatively, and the problem of a power-law
divergence is solved.

We now give an example for a precise definition of the
correlation function $C(x_0)$ and use the quenched approximation
to demonstrate that the continuum limit can be reached in all
steps while still a very interesting precision is attainable. 
The reader who is not interested in the numerical details may directly 
continue with \sect{s_fb}.
\subsubsection{Correlation functions, $\Oa$ improvement and 
               spin-symmetry \label{s_corr}}
In our numerical implementation we choose SF boundary conditions
with all details as in~\cite{zastat:pap3}, including $\theta=1/2$, $T=L$
and $C=C'=0$ in the notation of that paper. This means that 
$\Oa$ improvement~\cite{impr:pap1,zastat:pap1} is fully implemented, 
except for uncertainties in the coefficients $\castat$, $\ct$ and 
$\cttilde$ originating from their only perturbative estimation.
As in \cite{zastat:pap3} it has been checked that the influence of these
uncertainties on the observables considered here can be neglected
compared to our statistical errors. We 
will therefore not mention $\Oa$ terms any more and perform continuum
extrapolations modelling the $a$--effects as $\rmO(a^2)$.

\begin{figure}[htb]
\centering
\epsfig{file=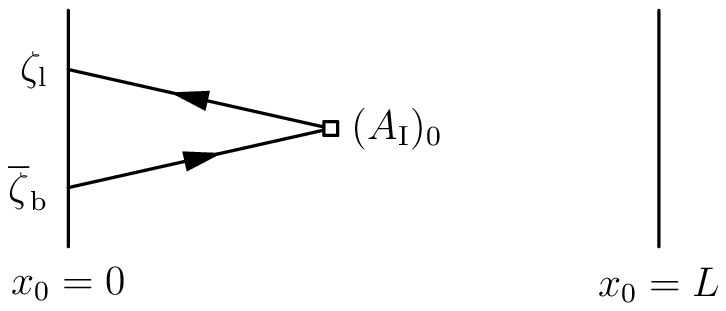,width=7.0cm}
\caption{ \footnotesize
Illustration of the correlation function $\fa$. 
For $\kv$, the insertion of $(\aimpr)_0$ is replaced by $(\vimpr)_k$,
whereas in case of $\fastat$ the operator in the bulk is $\Astatimpr$
connected to $\zetahb$ (instead of $\zetabar_{\rm b}$) by a static quark
propagator.
}\label{f_correlators}
\end{figure}

For our definition of $\meff$ we consider the two correlation functions
\bes
  \fa(x_0) &=& -{a^6 \over 2}\sum_{\vecy,\vecz}\,
  \left\langle 
  (\aimpr)_0(x)\,\zetabar_{\rm b}(\vecy)\gamma_5\zeta_{\rm l}(\vecz)
  \right\rangle  \,, \label{e_fa} \\
  \kv(x_0) &=& -{a^6 \over 6}\sum_{\vecy,\vecz,k}\,
  \left\langle 
  (\vimpr)_k(x)\,\zetabar_{\rm b}(\vecy)\gamma_k\zeta_{\rm l}(\vecz)
  \right\rangle  \,, \label{e_kv}
\ees 
where the label ``$\rm I$'' on the axial and vector currents reminds us 
that their $\Oa$ improved forms are used: 
\bes
  (\aimpr)_{\mu}(x) 
  & = &
  \lightb(x)\gamma_{\mu}\gamma_5\psi_{\rm b}(x)
  +a\ca\hlf(\drv{\mu}+\drvstar{\mu})\left\{
  \lightb(x)\gamma_5\psi_{\rm b}(x)\right\} \,,\\
  (\vimpr)_{\mu}(x) 
  & = &
  \lightb(x)\gamma_{\mu}\psi_{\rm b}(x)
  +a\cv\hlf(\drv{\nu}+\drvstar{\nu})\left\{
  i\,\lightb(x)\sigma_{\mu\nu}\psi_{\rm b}(x)\right\} \,.
\ees
As a consequence of the heavy quark spin-symmetry, their partners in the 
effective theory coincide exactly at lowest order in $\minv$ and we thus 
define only one 
\bes
 \label{e_fastat}
 C^{\rm HQET}(x_0) \equiv  \fastat(x_0) = 
 -{a^6 \over 2} \sum_{\vecy,\vecz}\,\left\langle 
 \Astatimpr(x)\,\zetahb(\vecy)\gamma_5\zeta_{\rm l}(\vecz)
 \right\rangle \,.
\ees
In these definitions,
$\sum_{\vecy,\vecz} \zetabar_{\rm b}(\vecy)\gamma_5\zeta_{\rm l}(\vecz)$
and $\sum_{\vecy,\vecz} \zetahb(\vecy)\gamma_5\zeta_{\rm l}(\vecz)$
are interpolating fields localized at the $x_0=0$ boundary of the SF,
which create a state with the quantum numbers of a B-meson with 
momentum $\vecp={\bf 0}$, and for $\gamma_5\to\gamma_k$ we
have the quantum numbers of a B$^{*}$.
The correlation functions are schematically depicted
in \fig{f_correlators}; more details can be found in \cite{zastat:pap1}.
%
%%%%%% figure: CF f_A
%

Inserting $C(x_0)=\fa(x_0)$ and $C(x_0)=\kv(x_0)$ into 
\eq{e_defgam} defines $\meffp$ and $\meffv$, respectively. Their
partner in the effective theory is denoted as $\meffstat$.
Due to the spin-symmetry, either $L\meffp=L(\meffstat+m)$ or 
$L\meffv=L(\meffstat+m)$ are possible matching conditions at lowest order 
in $\minv$. 
At first order, one has to consider also separately the
vector and the axial vector correlators in the effective theory
since these are split by the $\sigma\cdot\vecB$--term in the effective 
Lagrangian.
It is then convenient to define the matching condition as \eq{e_m1} with
the spin-average
\bes \label{e_spinav}
  \meff = \frac{1}{4}\left(\meffp + 3 \meffv\right) \,.
\ees
With this definition the matching condition (\ref{e_m1}) is independent
of the coefficient of the $\sigma\cdot\vecB$--term also at first order
in $\minv$, and hopefully $\minv$--effects are thereby minimized.

Having now completed our definition of $\meff$ and $\meffstat$,
we remind the reader that the mass of the light quark is set to zero
and thus $\meff$ is only a function of the heavy quark mass, $M$, and 
the linear extent of the SF-volume, $L$ (if $a$--effects are neglected
for the moment).
\subsubsection{Matching \label{s_omega}}
The essential steps of our strategy explained in \sect{s_strat} are the
matching at $L=L_0$ as the starting point, then connecting to $L=L_K$ 
and from there to the (infinite-volume) meson mass. 
In practice, proceeding from any choice of $u_0$ the sequence
$u_k$ is only known with a certain numerical precision and this
has to be taken into account in the error analysis. 
Furthermore we want to take advantage of 
the numerical results of \cite{mbar:pap1} for triples
of $(L/a,\beta,\kappa)$ corresponding to fixed renormalized coupling
and vanishing light quark mass in the quenched approximation, as 
well as of the known function $[r_0/a](\beta)$ and the value of 
$\Lmax/r_0=0.718(16)$ \cite{pot:r0_SU3}, where $\gbar^2(\Lmax)=3.48$. 
So it is convenient to have $K=2$ in previous formulae and to define 
(exactly) $L_2 = 2\Lmax = 1.436\,r_0$.
Then, applying the inverse of the step scaling function of the 
coupling \cite{mbar:pap1} twice, we arrive at
\bes \label{e_uvals}
  u_0 = 2.455(28)\,, \quad \sigma^{-1}(u_{0}) = 1.918(20) \,,
\ees
and end up with $L=L_0=L_2/4=0.359\,r_0\approx 0.18\,\fm$ for the linear
extent of the matching volume.
The matching of HQET and QCD is then supposed to be done at 
$ u = u_0 \approx 2.4$. We could thus keep e.g.~$u=2.4484$ fixed,
where the triples $(L/a,\beta,\kappa)$ are known \cite{mbar:pap1}, 
but only for $6 \leq L/a \leq 16$. The spacing of these lattices is still 
too large to comfortably accommodate a propagating b-quark. 
Instead it is better to work at a constant value of $u=1.8811$, 
varying $6 \leq L/(2a) \leq 16$. 
Nevertheless, $\Gamma$ is computed on the lattices with $L/a$ points per 
direction, and the slight mismatch of $\sigma^{-1}(u_{0})$ and $1.8811$ 
will eventually be taken into account together with the overall error 
analysis.

\begin{figure}[htb]
\centering
\epsfig{file=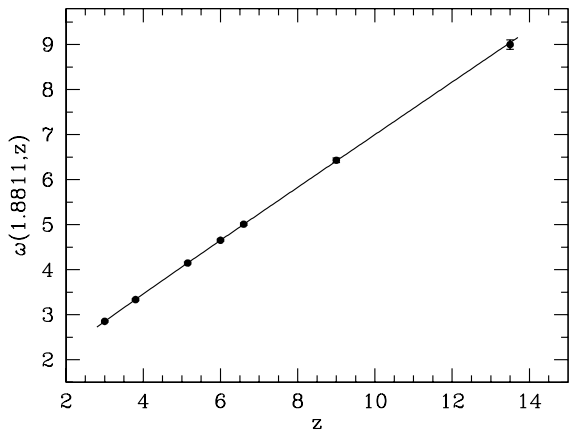,width=8.0cm}
\caption{{\footnotesize
Continuum limit values of $L_0\meff(L_0,M)$ at fixed coupling 
$\gbar^2(L_0/2)=1.8811$ as a function of $z=L_0M$ and its fit function, 
determined in relativistic QCD but small volume~\cite{mb:pap1}. 
The b-quark mass scale lies near $z\approx 6$.
}}\label{f_omega0}
\end{figure}

We now have to determine $\meff(L,M)$, where in case of the relativistic
theory $L$ is always to be identified with the extent of the matching 
volume, $L_0$, from now on. 
In order to approach its continuum limit, we define 
\bes
  \Omega(u,z,a/L) = \left. L\,\meff(L,M)\right|_{\gbar^2(L/2)=u\,,\,LM=z}
  \label{e_Omega}
\ees
and extrapolate it as a function of $a/L$, viz. 
\bes \label{e_omega0}
  \omega(u,z)=\lim_{a/L\to 0}\Omega(u,z,a/L)\,,
\ees
for a few selected values of $z$ and at fixed $u$. 
This requires to compute $\Omega$ with $z$ and $u$ fixed while changing 
$L/a$ and therefore also $\beta$.
A particular aspect in this step is that in imposing the condition of 
fixed $z$ (at variable $\beta$), the relation between the bare quark mass, 
$\mq$, and the RGI one, $M$, is needed, where several renormalization 
factors and improvement coefficients enter. 
Although they had already been determined \cite{mbar:pap1,impr:babp},
it turned out that it was desirable to improve their
precision and to determine them directly in the range of $\beta$ where they 
are needed  in the present application. 
For this reason they were redetermined in Ref.~\cite{mb:pap1}, and also 
$\omega$ as a function of $z$, \eq{e_omega0}, was obtained by extrapolation
in $a/L \to 0$ in that work. For the reader's convenience, 
we reproduce from~\cite{mb:pap1} a graph of the continuum values
$\omega(1.8811,z)$ together with the fit function
\bes
  \label{e_omegafit}
  \omega(1.8811,z) = a_0 z + a_1 + a_2\,\frac1z \,,\quad 
  a_0=0.581 \,,\quad
  a_1=1.226 \,,\quad
  a_2=-0.358
\ees
in \fig{f_omega0}.
%
%%%%%% figure: omega(1.8811,z)
%
In the interval $5.2 \le z \le 6.6$, which is the relevant $z$--range to
extract the RGI b-quark mass later, this parametrization describes 
$\omega(1.8811,z)$ with a precision of about 0.5\%.
A further global uncertainty of 0.9\% has to be attributed to the 
argument $z$ of the function $\omega$ (see Ref.~\cite{mb:pap1}).
In order to also take the small statistical error and mismatch 
in $u_0$ into consideration at the end, we also need a numerical value 
for the derivative of $\omega'(1.8811,z)$ w.r.t.~$u$. 
It was found to be  
constant in the interesting region \cite{mb:pap1}:
\bes
  \label{e_omegaprimefit}
  \left.{\partial \over \partial u}\,\omega(u,z)\right|_{u=1.8811} 
        = 0.70(1) \,, \quad 6.0\leq z\leq 6.6 \,.
\ees

For completeness we also quote the fit result for $\omega(1.8811,z)$,
if instead of the spin-average (\ref{e_spinav}) it is defined as the 
continuum limit of the effective energy $L\meffp(L,M)$.
With the same fit ansatz as 
in~\eq{e_omegafit}, the coefficients then read
\bes
  \label{e_omegafit_ps}  
  a_0=0.587  \,,\quad
  a_1=1.121  \,,\quad
  a_2=-1.306 \qquad
  \left(\mbox{for $\meff\equiv\meffp$}\right)\,.
\ees
The significantly larger $a_2$--term in this case compared 
to~\eq{e_omegafit} indicates that the spin-averaged
combination $\meff$ has smaller $\minv$--errors and should therefore be
preferred in the implementation of the matching step.\footnote{
In the preliminary computation of the b-quark's mass reported
in refs.~\cite{lat01:rainer,lat02:rainer}, the matching was performed
using only the energy in the pseudoscalar channel, $\meffp$. 
}
\subsubsection{Finite-size scaling \label{s_mb_fss}}
The next step is the numerical determination of the step scaling 
function $\sigmam$, \eq{e_sigmam}, and then of the 
($M$--independent difference)
$L_0[\meffstat(L_K,M)-\meffstat(L_0,M)]$ as given by \eq{e_m2}. 

%
%%%%%% figure: Sigma_m
%
\begin{figure}[htb]
\centering
\epsfig{file=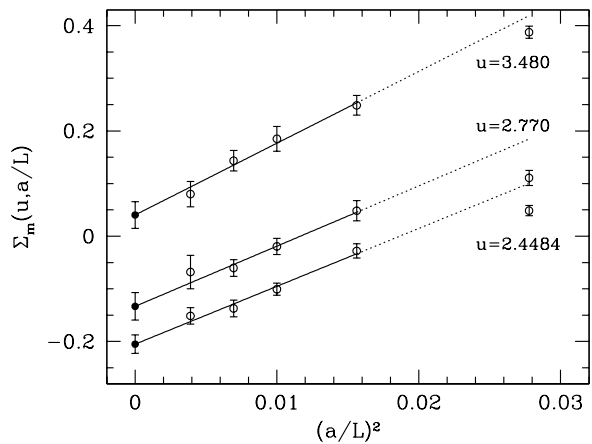,width=8.0cm}
\caption{{\footnotesize
Lattice step scaling function $\Sigma_{\rm m}$ and its continuum limit 
extrapolation linear in $(a/L)^2$, which uses 
only the four smallest values of $a/L$.
}}\label{f_contextr}
\end{figure}
\begin{figure}[htb]
\centering
\epsfig{file=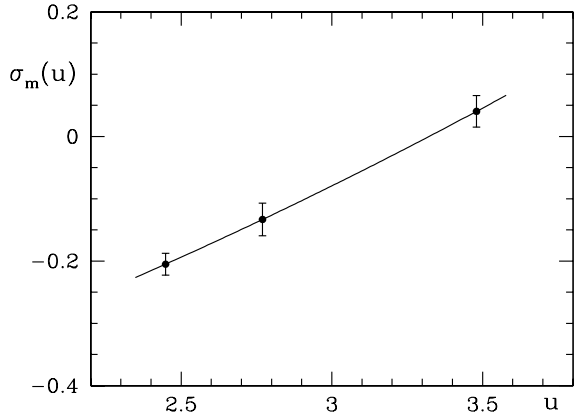,width=8.0cm}
\caption{{\footnotesize
Continuum step scaling function $\sigma_{\rm m}(u)$ and its polynomial fit.
}}\label{f_sigmam}
\end{figure}

%
%%%%%% table: sigma_m
%
\begin{table}[htb]
\centering
\begin{tabular}{cccr@{.}l}
\hline \\[-2.0ex]
  $u$ &&& \multicolumn{2}{c}{$\sigma_{\rm m}(u)$} \\[1.0ex]
\hline \\[-2.0ex]
  2.4484 &&& $-0$&$205(18)$                       \\    
  2.7700 &&& $-0$&$133(26)$                       \\
  3.4800 &&& $ 0$&$040(25)$                       \\[1.0ex]
\hline \\[-2.0ex]
\end{tabular}
\caption{{\footnotesize
Results for the continuum step scaling function $\sigma_{\rm m}(u)$.
}}\label{t_sigmam}
\end{table}
At finite lattice spacing, the step scaling function is defined by 
\bes
  \label{e_Sigmam}
  \Sigmam(u,a/L) =  
  \left. 2L\, 
  [\,\meffstat(2L,M) - \meffstat(L,M)\,]\right|_{\gbar^2(L)=u}\,,
\ees
where, as mentioned earlier, the (light) quark mass is set to zero. 
The exact definition of the massless point does
not play an important r\^ole; it is as in Ref.~\cite{zastat:pap3}.
In fact, we evaluated $\Sigmam$ directly from the correlation functions 
computed there, where details of the simulations can be found, too. 
Numerical results for $\Sigmam$ are listed in \tab{t_Sigmam} in the 
appendix. 
They are extrapolated to the continuum limit via
\bes
  \label{e_contextr}
  \Sigmam(u,a/L) = \sigmam(u) + c\, {a^2 \over L^2}\,.
\ees
Given our data at various resolutions, this is a safe extrapolation 
(cf.~\fig{f_contextr}) leading to continuum values reported 
in \tab{t_sigmam}.
Setting now $u_0$ as in \eq{e_uvals}, we want to compute 
(recall that $K=2$ now)
\bes \label{e_Deltameff}
  L_0\,[\,\meffstat(L_2,M)-\meffstat(L_0,M)\,] = 
  \frac12 \sigmam(u_0) + \frac14 \sigmam(u_1)\,,
\ees
with $u_1=\sigma(u_0)=3.48(5)$ \cite{mbar:pap1}. 
To this end it is convenient to represent the data of \tab{t_sigmam} by 
a smooth fit function,
\bes \label{e_sigmamfit}
  \sigmam(u) = s_0 + s_1 u +s_2 u^2\,,
\ees
shown in \fig{f_sigmam}.
%
%%%%%% figure: sigma_m
%
This fit implies the numerical value
\bes \label{e_res_sumssf}
  \frac12 \sigmam(u_0) + \frac14 \sigmam(u_1) = -0.092(11)
\ees
for the combination (\ref{e_Deltameff}). 
Here the uncertainties in $u_0,u_1$ are neglected, since
they would contribute only a negligible 
amount to the total error of \eq{e_res_sumssf}. 
Dropping the $s_2 u^2$--term in \eq{e_sigmamfit} would give
indistinguishable results.
\subsubsection{The subtracted B-meson mass \label{s_deltaE}}
As a last ingredient for the basic formula in \eq{e_mall} we have to
address $L_0\Delta E$, with $\Delta E$ the energy difference between the
B-meson's static binding energy and the effective energy of the 
static-light correlator $\fastat$, \eq{e_m3}, at the same values of the
lattice spacing.
We evaluate this quantity starting
from results for $a\Estat(g_0)$ in the literature \cite{stat:fnal2}. 
They are interpolated in the mass of the light quark
to the strange quark mass (see also Sect.~5.2.~of Ref.~\cite{zastat:pap3}) 
and then correspond to a ${\rm B}_{\rm s}$--state. Since $\Oa$ improvement 
was not employed in the computation of Ref.~\cite{stat:fnal2}, we also 
need $\meffstat(L_2)$ for the unimproved theory. 
Given the simulation results reported in Appendix~C.2.~of 
Ref.~\cite{zastat:pap3}, $a\meffstat(g_0,L/a)$ with $L=L_2=1.436\,r_0$ is 
straightforwardly obtained for $5.68 \le \beta=6/g_0^2\le 6.5$.
These numbers are collected in \tab{t_gamstat_1.436r0} of the appendix and 
well described by
\bes
\left. a\meffstat(g_0,L/a) \right|_{L=L_2}=
0.394-0.055\,(\beta-6)-0.218\,(\beta-6)^2+0.229\,(\beta-6)^3
\label{res_gamstat1.436r0wil}
\ees
with an absolute uncertainty of less than $\pm 0.002$ in the range of 
$\beta$ mentioned.

\begin{figure}[htb]
\centering
\epsfig{file=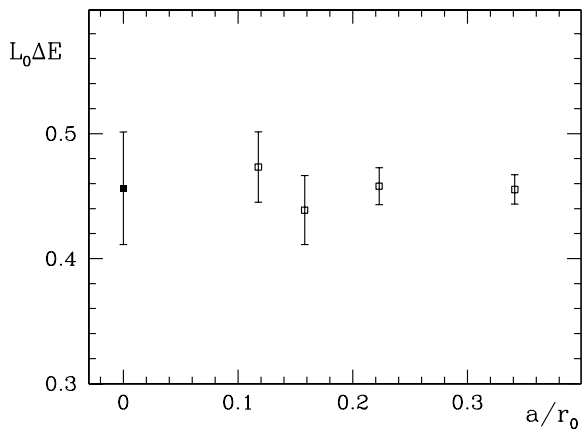,width=8.0cm}
\caption{{\footnotesize
Subtracted, dimensionless ${\rm B}_{\rm s}$--meson energy evaluated from 
the bare numbers of \protect\cite{stat:fnal2}.
}}\label{f_DeltaE}
\end{figure}

The combination $L_0\Delta E$ is shown in \fig{f_DeltaE}. 
%
%%%%%% figure: L_0*Delta_E
%
Its errors are dominated by those of $a\Estat$.  
Since they are rather large and also the lattice spacings are not very 
small, we refrain from forming a continuum extrapolation. 
Instead we take
\bes
  L_0\Delta E = 0.46(5)
\ees
as our present estimate. 
As seen in the figure, its error encompasses the full range of results at 
finite $a$ and the true continuum number is expected to be covered by it.
 
No doubt, a continuum limit with small error (at least a factor 3 smaller)
can be achieved here in the near future, incorporating $\Oa$ improvement 
and using the alternative discretization of static quarks of 
Ref.~\cite{stat:letter}.
\subsubsection{Determination of the quark mass}
\begin{figure}[htb]
\centering
\epsfig{file=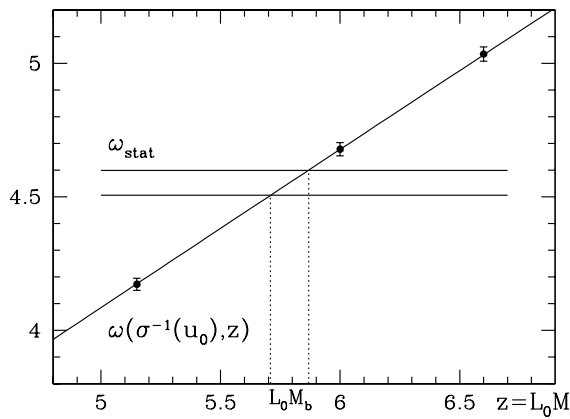,width=8.0cm}
\caption{{\footnotesize
Graphical solution of the basic formula, \eq{e_mall}, for the dimensionless
RGI b-quark mass, $z_{\rm b}\equiv L_0\Mbeauty$. 
The contributing pieces are repeated explicitly 
in eqs.~(\ref{e_omegashift}) and (\ref{e_cmb2solve}).
}}\label{f_solveMb}
\end{figure}

Now we are in the position to put all pieces together and solve the basic
equation (\ref{e_mall}) for the b-quark mass.
This amounts to determine the interception point of the function 
\bes
  \omega\left(\sigma^{-1}(u_0),z\right) = 
  \omega(1.8811,z) 
  + \Delta u \left.{\partial \over \partial u}\,\omega(u,z)
  \right|_{u=1.8811} \,,
  \label{e_omegashift}
\ees
where $\Delta u=(\sigma^{-1}(u_0)-1.8811)=0.037(20)$ accounts for the 
slight mismatch in the couplings fixed, with the combination
\bes
  \omega_{\rm stat} \equiv
  L_0\mb
  - \left\{\,   \textstyle{\frac{1}{2}}\,\sigmam(u_0)
              + \textstyle{\frac{1}{4}}\,\sigmam(u_1) \,\right\}
  - L_0\Delta E \,.
  \label{e_cmb2solve}
\ees
All of the necessary quantities are known from the foregoing three 
subsections and the experimental spin-averaged B-meson mass 
$\mb=m_{{\rm B}_{\rm s}}=5.40\,\GeV$ is taken as the physical input.
As illustrated in \fig{f_solveMb}, this procedure yields a value for 
$L_0\Mbeauty$ with an error. 
%
%%%%%% figure: graphical solution for M_b
%
Together with $L_0/r_0=\Lmax/(2r_0)=0.359$ it is converted to our central 
result
\bes \label{e_resMb}
  r_0 \Mbeauty = 16.12(25)(15) \,.
\ees
Here the second error in parentheses stems from the additional 
0.9\% uncertainty of $z$ in $\omega(u,z)$ that was mentioned in the
context of \eq{e_omegafit}.
With $r_0=0.5\,\Fm$ this translates into 
\bes
 \Mbeauty = 6.36(10)(6)\,\GeV \,,\quad 
 \mbar_{\rm b}\big(\mbar_{\rm b}\big) = 4.12(7)(4)\,\GeV \,,
\ees
where the running mass, $\mbar_{\rm b}(\mu)$, is evaluated with the 
four-loop renormalization group functions and in the $\MS$ scheme. 

One should remember that this result is valid up to 
$\minv$--corrections and it has been obtained in the 
quenched approximation. One may estimate the ambiguity due to setting the
energy scale in the quenched approximation as usual.
Varying the value of $r_0$ in fm by 10\%, we obtain changes of about 
$260\,\MeV$ in $\Mbeauty$ and $150\,\MeV$ in 
$\mbar_{\rm b}(\mbar_{\rm b})$. 
We emphasize, however, that the scale ambiguity can serve only as a rough 
guide to the impact of quenching.
Finally, we also want to stress once more that in \sect{s_deltaE} the 
contribution from the subtracted B-meson mass to our result on the b-quark 
mass is still based on ordinary Wilson fermion data and will be hopefully
much improved soon.
\subsection{The B-meson decay constant at lowest order \label{s_fb}}
To lowest order in $\minv$, the bare matrix element
\bes
  \calM(g_0) = \langle {\rm B}(\vecp={\bf 0})| \Astat(0) |0 \rangle
\ees
evaluated with the static action together with a renormalization factor
$\zastat(g_0,a\Mb)$ allows to determine the B-meson decay constant
in static approximation~\cite{stat:eichten}:
\bes \label{e_fb}
  \Fb\sqrt{\mb} = \lim_{a\to 0}  \zastat(g_0,a\Mb) \calM(g_0)\,. 
\ees
The renormalization constant $\zastat(g_0,a\Mb)$ has already been computed
in the quenched approximation \cite{zastat:pap3}, using a quite different
method to the one described in this paper. 
However, the method of Ref.~\cite{zastat:pap3} is not easily
extended to include $\minv$--corrections. Therefore, as a second example of 
the use of our general strategy, 
we here describe an alternative method which may be 
extended to include $\minv$--corrections.

The key formula, valid up to corrections of $\rmO(\minv)$, has already 
appeared and was briefly discussed in Ref.~\cite{lat02:rainer}:
\bes \label{e_fbfactors}
  \Fb\sqrt{\mb} = { \left. \Fb\sqrt{\mb}\right|^{\rm HQET}
                  \over \Phihqet(L_2,\Mb) }
                  \times{ \Phihqet(L_2,\Mb)  \over \Phihqet(L_1,\Mb) }
                  \times{ \Phihqet(L_1,\Mb)  \over \Phihqet(L_0,\Mb) }
                  \times  \Phiqcd(L_0,\Mb) \,. \nonumber\\[-1ex]
\ees
In the rest of this section
we give precise definitions of the various factors and
explain the formula in the general framework of 
Sects.~\ref{s_match} and \ref{s_finvol}.
We assume that the b-quark mass is already known. 
Terms necessary for $\Oa$ improvement are not written explicitly; 
they can easily be added.
\subsubsection{Matching}
As a preparation for the matching of the axial current and the associated
determination of $\zastat\equiv \acoeff{0}{0}$,
we start from
$\fastat(x_0)$, defined in \sect{s_corr}. This correlation
function renormalizes multiplicatively:
$\fastatren = Z_\zetah Z_\zetal \zastat \fastat$.  
In order to eliminate the 
renormalization factors $Z_\zetah, Z_\zetal$  of the boundary quark fields,
we consider in addition the boundary-to-boundary correlation,
\bes
  \fonestat = 
  -{a^{12} \over 2L^6}\sum_{\vecu,\vecv,\vecy,\vecz}
  \left\langle
  \zetalbprime(\vecu)\gamma_5\zetahprime(\vecv)\,
  \zetahb(\vecy)\gamma_5\zetal(\vecz)
  \right\rangle\,,           
\ees
which is renormalized as $\fonestatren= (Z_\zetah Z_\zetal)^2 \fonestat$.
In the ratio
\bes \label{e_X}
  X(g_0,L/a)\equiv{\fastat(L/2) \over \sqrt{\fonestat}}
\ees 
the unwanted $Z$--factors cancel, and with the choice $x_0=L/2$ it
is also independent of the linearly divergent mass counterterm $\dmstat$.
The renormalized ratio is denoted by
\bes \label{e_phihqetfb}
  \Phihqet(L,\Mb) 
  & \equiv & 
  \xr(u,\zb,a/L) \\ \nonumber
  &    =   &
  \left.\zastat(g_0,a\Mb)\,X(g_0,L/a)\right|_{\gbar^2(L)=u}\,,\quad
  \zb=L\Mb \,.
\ees
Here, as wherever we do not mention the light quark masses explicitly, 
they are set to zero. 

In QCD we define the corresponding quantities ($\za$ denotes the 
standard renormalization constant for the QCD axial current 
\cite{Boch,impr:pap4} and $\fone$ is the analogue of $\fonestat$ with
$\zeta_{\rm h}\to\zeta_{\rm b}$)
\bes
  Y(g_0,\zb,L/a)&=&{\fa(L/2) \over \sqrt{\fone}}\,,\\
  \yr(u,\zb,a/L)&=&\left.\za(g_0)\,Y(g_0,\zb,L/a)\right|_{\gbar^2(L)=u}\,,\\
  \Phiqcd(L,\Mb)&=& \lim_{a/L\to 0}\yr(u,\zb,a/L)\,,
  \label{e_phiqcdfb}
\ees
and the matching equation to be imposed in the small volume (of extent
$L_0$) is 
\bes \label{e_mphi} 
  \Phihqet(L_0,\Mb) = \Phiqcd(L_0,\Mb) \quad {\rm with} \quad 
  \gbar^2(L_0)=u_0={\rm fixed}\,.
\ees
While in agreement with the notation in previous sections $\Phihqet$ on the 
l.h.s.~has a lattice spacing dependence that is only implicit
(cf.~\eq{e_phihqetfb}), for the r.h.s.~the continuum limit is to be taken 
(cf.~\eq{e_phiqcdfb}). 
In the quenched approximation, the particular value for $u_0$ may be chosen 
as in \sect{s_mb}.
\subsubsection{Finite-size scaling}
Computing the step scaling functions built as 
\bes \label{e_sigmax}
  \sigma_{\rm X}(u) \equiv 
                    \lim_{a/L \to 0} 
                    \left.{ X(g_0,2L/a) \over X(g_0,L/a)} 
                    \right|_{\gbar^2(L)=u}
\ees
allows to reach larger values of $L$ via the recursion
\bes \label{e_recursion_x}
  \left.\Phihqet(2L,\Mb)\right|_{a=0} = 
  \sigma_{\rm X}\left(\gbar^2(L)\right) 
  \times \left.\Phihqet(L,\Mb)\right|_{a=0} \,.
\ees 
We note in passing that in \eq{e_sigmax} we could have written $X\to \xr$,
since the same $\zastat$ enters in numerator and denominator. 
For the following we assume \eq{e_recursion_x} to be iterated 
$K$ times, connecting
$\Phihqet(L_0,\Mb)$ to $\Phihqet(L_K,\Mb)$ with $L_K=2^K L_0$,
i.e.~in a numerical implementation, details will be very similar to those 
described in \sect{s_mb_fss}.
\subsubsection{The decay constant}
To finally arrive at $\Fb\sqrt{\mb}$, one defines the renormalization
constant in view of \eq{e_phihqetfb},
\bes \label{e_zafin}
\zastat(g_0,a\Mb) = { \xr(u_K,\zb,0) \over X(g_0,L_K/a) } 
                  = { \Phihqet(L_K,\Mb)|_{a=0} \over X(g_0,L_K/a)}\,,
\quad u_K=\gbar^2(L_K) \,,
\ees
and --- bearing in mind that $\Phihqet(L_K,\Mb)|_{a=0}$ via 
\eq{e_recursion_x} and the matching condition (\ref{e_mphi}) may be evolved 
back to the \emph{renormalized} quantity $\Phiqcd(L_0,\Mb)$ in QCD --- 
replaces $\zastat(g_0,a\Mb)$ in \eq{e_fb}  by rewriting:
\bes
  \Fb\sqrt{\mb}&=&\rho(u_K)\times\left.\Phihqet(L_K,\Mb)\right|_{a=0} \,,\\
       \rho(u) &=&\lim_{a/L\to 0} R(u,a/L)\,,\quad 
  R(u,a/L)\equiv\left.{\calM(g_0)\over X(g_0,L/a)}\right|_{\gbar^2(L)=u}\,.
\ees
Although we have chosen only $u$ and $a/L$ as arguments for $R$,
it does depend on the masses of the light quarks
used for the evaluation of the matrix element $\calM$. 
These masses have to be set
(or extrapolated) to the physical ones to obtain the correct matrix
element in question; conveniently, the light quark masses are put to zero
in all other quantities as mentioned above. In this way the effective theory 
is renormalized in a (light quark) mass-independent renormalization scheme.

Choosing as an example $K=2$, we may combine all ingredients
into the expression
\bes
  \Fb\sqrt{\mb} = \rho(u_2)\times \sigma_{\rm X}(u_1)\times
                  \sigma_{\rm X}(u_0)  
                  \times \Phiqcd(L_0,\Mb)\,,
\ees
where the various factors correspond one to one to those in
\eq{e_fbfactors} but are now rigorously defined and can be computed in the
continuum limit. 
Our notation is most appropriate for the lowest order in $\minv$. 
At higher order, additional matching equations and step scaling functions
have to be defined and $\sigma_{\rm X}$ will depend on $\Mb$,
which is not the case at lowest order in $\minv$. 
In fact, at this order all the (heavy quark) mass dependence of 
$\Fb\sqrt{\mb}$ is contained in $\Phiqcd(L_0,\Mb)$ that is calculable in
small-volume lattice QCD with a relativistic b-quark \cite{hqet:pap2}. 
\subsubsection{Relation to other approaches \label{s_relation}}
We finally want to compare the strategy of renormalizing 
$\Astat$, presented in this paper, to the one of 
\cite{zastat:pap1,zastat:pap3}, which followed the general
ALPHA Collaboration strategy of obtaining the renormalization
group invariant matrix elements of composite operators 
non-perturbatively and then using (continuum) perturbative results to 
find the operators normalized in the matching scheme at finite 
renormalization scale.
(A recent review is Ref.~\cite{lat02:rainer}.) 
For the application to the HQET, the natural renormalization scale is 
then the mass of the b-quark.

Remaining strictly at lowest order in
$\minv$, there is no mixing with 
lower-dimensional operators. Consequently, perturbation theory
can be applied. In particular the large-mass 
behaviour of $\Fb\sqrt{\mb}$, which is given by the mass
dependence of $\zastat$, is computable leading to the 
asymptotics \cite{ShifVol,PolWise}\footnote{
The leading-order coefficient $b_0$ of the $\beta$--function is taken for 
$\nf-1$ flavours, since the b-quark does not contribute.
Note that here $\nf-1=0$ corresponds to the quenched approximation.}
\bes \label{e_rgi}
  \lim_{\Mb\to\infty} 
  \left[\,\ln(\Mb/\Lambda_\msbar)\,\right]^{\gamma_0/(2b_0)}\Fb\sqrt{\mb}
  &=& 
  \FhatstatRGI \,\,\,=\,\,\, {\rm constant}\,, \\ \nonumber
  & & 
  \gamma_0=-\frac{1}{ 4\pi^2}\,, \quad
  b_0=\frac{11-\frac{2}{3}(\nf-1)}{ 16 \pi^2} \,. 
\ees
Furthermore, the function 
$\Cps(\Mb/\Lambda_\msbar) = \Fb\sqrt{\mb}/\FhatstatRGI$  
is known perturbatively up to and including 
$\gbar^4(\mbar_{\rm b})$--corrections to the leading-order 
equation (\ref{e_rgi}) 
\cite{Ji:1991pr,BroadhGrozin,Gimenez:1992bf,BroadhGrozinII,Grozin,ChetGrozin}. 
These perturbative computations, like our non-perturbative method, are 
based on the renormalization of the static axial current where the
finite part is determined by matching, \eq{e_match}. 
We denote this renormalization scheme by the ``matching scheme'' 
\cite{lat02:rainer,zastat:pap3}.\footnote{
Non-perturbatively, the matching scheme is unique up
to higher orders in $\minv$; in perturbation theory, a residual 
scheme dependence due to the choice of the other renormalized
parameters, such as $\gbar$, remains.}

A finite, renormalized static axial current can of course also be 
defined by other renormalization conditions involving
a renormalization scale $\mu$ in a suitably chosen intermediate
scheme. Matrix elements $\Phiinter(\mu)$
of the renormalized current in this scheme will
then not necessarily satisfy \eq{e_equiv}, but it is easy to see
that 
\bes \label{e_rgiinter}
  \lim_{\mu\to\infty} 
  \left[\,8\pi b_0 \alpha(\mu)\,\right]^{-\gamma_0/(2b_0)}\Phiinter(\mu)  
  =\FhatstatRGI \,,\quad \alpha=\gbar^2/(4\pi)\,,
\ees
is independent of the intermediate renormalization scheme.
In Ref.~\cite{zastat:pap3}
a finite-volume scheme was adopted,
which allows to evaluate the limit in \eq{e_rgiinter} through some
finite-size scaling steps followed by perturbative 
evolution at very high $\mu$. The results obtained in this
way are then combined with the perturbative approximation
of $\Cps(\Mbeauty/\Lambda_\msbar)$. 
Owing to this last step, they are accurate up to relative errors of order 
$\gbar^6(\mbar_{\rm b})$.
This discussion should have made evident that the essential difference of 
the method presented here is not the absence of these perturbative errors, 
which are expected to be quite small, but rather
the tempting possibility to {\em include $\minv$--corrections}.

\section{Uncertainties and perspectives \label{s_uncert}}
Following the general strategy introduced in this paper opens the possibility
to perform clean non-perturbative computations using the lattice regularized
HQET. The dangerous power-law divergences are subtracted non-perturbatively
through the matching in small volume. This is not only a theoretical 
proposal: in \sect{s_mb} we showed that in a concrete physics application
the statistical errors of Monte Carlo results are quite moderate. 
In fact they can be expected to be even smaller, when an alternative
discretization of the static approximation is employed \cite{stat:letter}.

We emphasize that the result for the b-quark mass in \sect{s_mb}
is valid up to $1/\Mbeauty$--corrections. If we had used $\meff=\meffp$ 
instead of the spin-averaged energy in the matching step, a $0.4\,\GeV$ 
higher number for $\Mbeauty$ would
have been obtained. We do not regard this shift as a realistic
estimate of the magnitude of $1/\Mbeauty$--corrections but believe that they 
are significantly smaller, as indicated by the smallness of the associated
coefficient $a_2$ in the numerically determined quark mass dependence of
the spin-averaged energy, \eq{e_omegafit}. 
Nevertheless it is clear that a precision 
determination of $\Mbeauty$ requires to take the $1/\Mbeauty$--corrections 
into account, even if only to show that they are small.

In general one may argue that the matching step 
(carried out at order $1/\Mbeauty^n$) contains 
$1/(L_0 \Mbeauty)^{n+1}$--uncertainties in addition to the unavoidable
$1/(r_0 \Mbeauty)^{n+1}$--terms (we remind the reader that we take
$1/r_0 \approx 0.4\,\GeV$ as a typical QCD scale). 
Whether or not the former terms are larger 
than the latter can only be decided if the linear extent of the matching
volume, $L_0$, is varied. Increasing it
would demand even smaller lattice spacings.  

Since $1/\Mb$--corrections are computable, they should be determined 
to arrive at precision computations for B-physics observables, 
e.g.~for the phenomenologically interesting B--$\overline{\rm B}$ mixing
amplitude. On the other hand
we expect $(1/\Mb)^2$--corrections to be very difficult in practice, 
because they require also O($a^2$) improvement of the whole theory.
Fortunately, $(1/\Mb)^2$--corrections do not appear to be very 
important~\cite{hqet:pap2} but, as in every expansion, 
this issue has to be studied case by case.

An attractive property of our strategy 
is that it does not involve any particularly large lattices and
therefore all the steps outlined in the present work can also be performed 
with dynamical fermions. These calculations are presumably no more
difficult than for instance those of 
Refs.~\cite{spect:cppacs,UKQCD:nf2b52,JLQCD:nf2b52}
in the light meson sector.
\subsection*{Acknowledgements}
We would like to thank M.~Della Morte, A.~Shindler and U.~Wolff
for useful discussions and comments on the manuscript. This work
is part of the ALPHA Collaboration research plan. 
We thank DESY for allocating computer time on the APEmille computers
at DESY Zeuthen to this project and the APE-group for its valuable
help. 
We further acknowledge support by the EU IHP Network on
{\em Hadron Phenomenology from Lattice QCD} under grant HPRN-CT-2000-00145
and by the Deutsche Forschungsgemeinschaft in the SFB/TR 09.

\begin{appendix}
\section{Detailed numerical results\label{s_numres}}
In this appendix we collect some of the  numerical results underlying 
the computation of the b-quark's mass in \sect{s_mb}.

%
%%%%%% table: Sigma_m
%
\begin{table}[htb]
\centering
\begin{tabular}{lccr@{.}lrr@{.}lr@{.}lr@{.}l}
\hline
  $\gbar^2(L)$ && $\beta=6/g_0^2$ & \multicolumn{2}{c}{$\kappa$} & $L/a$ 
& \multicolumn{2}{c}{$a\meffstat(L)$} 
& \multicolumn{2}{c}{$a\meffstat(2L)$} 
& \multicolumn{2}{c}{$\Sigma_{\rm m}(u,a/L)$} \\
\hline
  2.4484 &&
  6.7807 & 0&134994 &  6 & $0$&$3082(6 )$ & $0$&$3123(6 )$ & $ 0$&$049(10)$
\\
         &&
  7.0197 & 0&134639 &  8 & $0$&$2899(4 )$ & $0$&$2881(7 )$ & $-0$&$028(13)$
\\
         &&
  7.2025 & 0&134380 & 10 & $0$&$2747(4 )$ & $0$&$2697(4 )$ & $-0$&$101(12)$
\\
         &&
  7.3551 & 0&134141 & 12 & $0$&$2621(5 )$ & $0$&$2564(5 )$ & $-0$&$137(16)$
\\
         &&
  7.6101 & 0&133729 & 16 & $0$&$2444(2 )$ & $0$&$2396(4 )$ & $-0$&$151(16)$
\\
\hline
  2.7700 &&
  6.5512 & 0&135327 &  6 & $0$&$3282(6 )$ & $0$&$3374(10)$ & $ 0$&$111(14)$
\\
         &&
  6.7860 & 0&135056 &  8 & $0$&$3068(8 )$ & $0$&$3098(9 )$ & $ 0$&$048(19)$
\\
         &&
  6.9720 & 0&134770 & 10 & $0$&$2916(4 )$ & $0$&$2907(6 )$ & $-0$&$019(15)$
\\
         &&
  7.1190 & 0&134513 & 12 & $0$&$2795(4 )$ & $0$&$2769(5 )$ & $-0$&$060(16)$
\\
         &&
  7.3686 & 0&134114 & 16 & $0$&$2592(6 )$ & $0$&$2570(8 )$ & $-0$&$068(32)$
\\
\hline
  3.4800 &&
  6.2204 & 0&135470 &  6 & $0$&$3629(6 )$ & $0$&$3952(7 )$ & $ 0$&$388(12)$
\\
         &&
  6.4527 & 0&135543 &  8 & $0$&$3399(5 )$ & $0$&$3554(11)$ & $ 0$&$249(19)$
\\
         &&
  6.6350 & 0&135340 & 10 & $0$&$3219(6 )$ & $0$&$3311(10)$ & $ 0$&$185(23)$
\\
         &&
  6.7750 & 0&135121 & 12 & $0$&$3081(3 )$ & $0$&$3141(7 )$ & $ 0$&$143(20)$
\\
         &&
  7.0203 & 0&134707 & 16 & $0$&$2858(3 )$ & $0$&$2883(7 )$ & $ 0$&$080(24)$
\\
\hline \\[-2.0ex]
\end{tabular}
\caption{{ \footnotesize
Results for the lattice step scaling function $\Sigma_{\rm m}$.
For the intermediate quantity $\meffstat(L)$ the argument $M$ is suppressed,
since it is evaluated at $\mhbare=0$ and so does not depend on $M$.
}}\label{t_Sigmam}
\end{table}
The numerical data on the static-light meson correlator in the Schr\"odinger 
functional (SF), which are necessary to evaluate its logarithmic derivative 
$\meffstat$, see eqs.~(\ref{e_defgam}) and (\ref{e_fastat}), have already 
been obtained in the context of the non-perturbative renormalization of the 
static-light axial current, Ref.~\cite{zastat:pap3}.
Hence we refer the reader to this work for more details.
In \tab{t_Sigmam} we list the numerical results on the lattice step scaling
function $\Sigma_{\rm m}$, defined through \eq{e_Sigmam}.

\begin{figure}[htb]
\centering
\epsfig{file=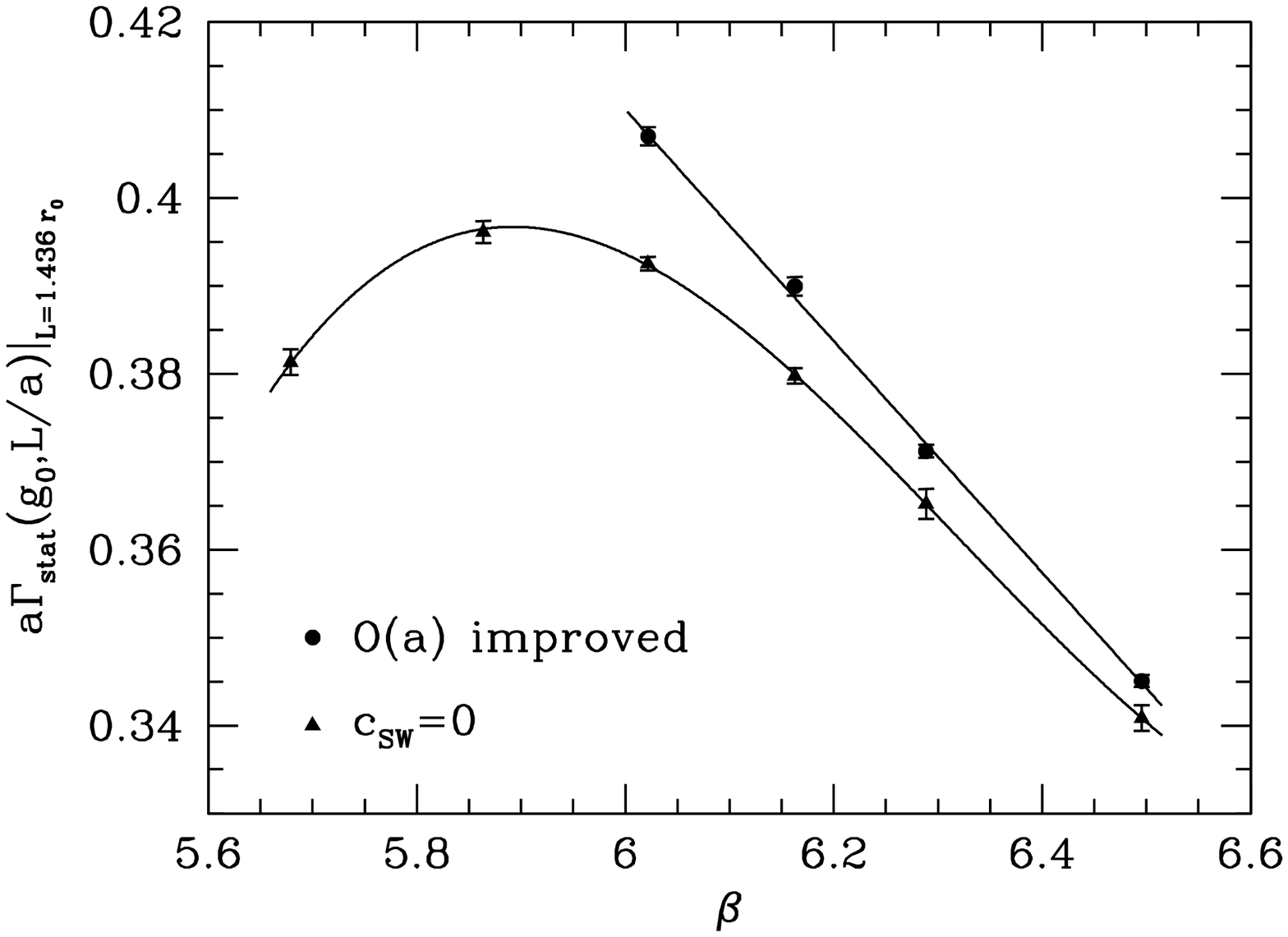,width=8.0cm}
\caption{{ \footnotesize
Numbers for $a\meffstat(g_0,L/a)|_{L=1.436\,r_0}$ from
\tab{t_gamstat_1.436r0} and its polynomial fit functions for $\Oa$ 
improved and unimproved  Wilson fermions.
}}\label{f_gamstat_1.436r0}
\end{figure}

Another ingredient in the determination of $\Mbeauty$ is the subtracted
B-meson energy $\Delta E$, \eq{e_m3}, which amounts to calculate the static 
effective energy $\meffstat$ at
$L=1.436\,r_0$.
Our quenched results for this quantity, both for the $\Oa$ improved case 
(with non-perturbative $\csw$ from \cite{impr:pap3} and the 
1--loop value \cite{MORNINGSTAR1} for the coefficient $\castat$ to improve 
the static-light axial current) as well as for the unimproved theory
(where both improvement coefficients are set to zero), are shown 
in \tab{t_gamstat_1.436r0} and \fig{f_gamstat_1.436r0}. 
These numbers are well represented by the polynomial expressions
\[
\left. a\meffstat(g_0,L/a)\right|_{L=1.436\,r_0}=
\left\{
\begin{array}{l}
0.410-0.132\,(\beta-6)\\
\mbox{for $6.0 \le \beta=6/g_0^2\le 6.5$ 
      and $\csw=\mbox{non-perturbative}$}\\
\\
0.394-0.055\,(\beta-6)-0.218\,(\beta-6)^2+0.229\,(\beta-6)^3\\
\mbox{for $5.68 \le \beta=6/g_0^2\le 6.5$ and $\csw=0$}
\end{array}
\right.\,,
\]
where their absolute uncertainty is below $\pm 0.001$ and $\pm 0.002$ in 
the indicated ranges of $\beta$, respectively.
In \sect{s_deltaE} we restrict the analysis only to the case of unimproved 
Wilson fermions, $\csw=0$, but the $\Oa$ improved parametrization may 
be required when also $\Oa$ improved data on the static binding 
energy $\Estat$ at various lattice spacings will become available.
%
%%%%%% table: a*Gamma_stat(g_0,L/a)_{L=2*L_max} (O(a)- and unimproved)
%
\begin{table}[htb]
\centering
\begin{tabular}{rccccccc}
\hline \\[-2.0ex]
&&& 
\multicolumn{2}{c}{$\Oa$ improved} 
&& 
\multicolumn{2}{c}{$\csw=0$} \\[1.0ex]
  $L/a$ & $\beta=6/g_0^2$ &&  
  $\kappa$ & $a\meffstat$ & & $\kappa$ & $a\meffstat$     \\[1.0ex]
\hline \\[-2.0ex]
  4  & 5.6791 && ---     & ---      && 0.15268 & 0.381(2) \\
  6  & 5.8636 && ---     & ---      && 0.15451 & 0.396(1) \\
  8  & 6.0219 && 0.13508 & 0.407(1) && 0.15341 & 0.393(1) \\
  10 & 6.1628 && 0.13565 & 0.390(1) && 0.15202 & 0.380(1) \\
  12 & 6.2885 && 0.13575 & 0.371(1) && 0.15078 & 0.365(2) \\
  16 & 6.4956 && 0.13559 & 0.345(1) && 0.14887 & 0.341(2) \\[1.0ex]
\hline \\[-2.0ex]
\end{tabular}
\caption{{ \footnotesize
Numerical results for $a\meffstat$ at $L=1.436\,r_0$ for $\Oa$ improved 
and unimproved Wilson fermions 
(i.e.~$\csw=0$ and also $\castat=0$ in the latter case).
The corresponding simulation parameters are reproduced from 
Appendix~C.2.~of \cite{zastat:pap3} for completeness.
}}\label{t_gamstat_1.436r0}
\end{table}
%
%%%%%% figure: a*Gamma_stat(g_0,L/a)_{L=2*L_max} (O(a)- and unimproved)
%

\end{appendix}
\bibliography{lattice}
\bibliographystyle{h-elsevier3}
\end{document}